%Paper: hep-th/9301072
%From: gamboa@ipncls.in2p3.fr
%Date: Mon, 18 Jan 93 09:41:41 GMT

\magnification=1200
\def\dsl{\raise.15ex\hbox{/}\kern-.57em\partial}
\def\Dsl{\,\raise.15ex\hbox{/}\mkern-13.5mu D} %this one can be subscripted
\def\Asl{\,\raise.15ex\hbox{/}\mkern-13.5mu A} %this one can be subscripted
\def\Bsl{\,\raise.15ex\hbox{/}\mkern-13.5mu B} %this one can be subscripted
\def\Qsl{\,\raise.15ex\hbox{/}\mkern-13.5mu \Tr} %this one can be subscripted
\def\Ksl{ \overleftarrow}
\def\Sn{\sum_{\bf n}}
\def\nf{\bf n}
\def\Tr{{\triangle}}
\def\cc{\overrightarrow}
\font\title=cmr12 at 12pt
\footline={\ifnum\pageno=1\hfill\else\hfill\rm\folio\hfill\fi}
\baselineskip=18pt
\rightline{DFTUZ 92/14}
\rightline{IPNO-TH 92/83}
\rightline{September 1992}
\vskip 1.0cm
\centerline{\title SECOND ORDER FORMALISM FOR FERMIONS:}
\centerline{\title ANOMALIES AND SPECIES DOUBLING}
\centerline{{\title ON THE LATTICE}
\footnote{$^\dagger$}{Work partially supported by CICYT (Proyecto AEN
90-0030).}}
\vskip 2.0cm
\centerline{\bf J. L. Cort\'es$^1$, J. Gamboa$^2$ and
L. Vel\'azquez$^1$}
\centerline{\it $^1$Departamento de F\'\i sica Te\'orica, Universidad de
Zaragoza,}  \centerline{\it 50009 Zaragoza, Spain.}
\centerline{\it $^2$Divisi\'on de Physique The\'orique, Institut de Physique
Nucleaire\footnote{$^\ddagger$}{\it Unite de Recherche des Universit\'es
Paris 11 et Paris 6 associe au CNRS},}
\centerline{\it F-91406 Cedex, Orsay, France.}
 \vskip 1.0cm
{\bf Abstract}. A new formulation for fermions on the lattice based on a
discretization  of a second order formalism is proposed. A comparison with the
first order  formalism in connection with the $U(1)$ anomaly and the doubling
problem  is presented. The new formulation allows to eliminate half of the
degrees  of freedom, which can be important in non-perturbative
calculations, and  it can open new possibities to the formulation of chiral
gauge
theories. \vskip 2.0cm PACS 11.15.Ha , 11.30.Rd  \vfill \eject

Many applications of relativistic quantum field theory in particle physics
involve as an important ingredient a non-perturbative treatment of a fermionic
degree of freedom, for which only limited results have been obtained. In a
formulation based on a discretization of spacetime, it is only after a
considerable progress in computing capabilities that, simulations going beyond
the quenched approximations have been possible, although still with serious
limitations.
This makes interesting any attempt to simplify the treatment of a fermionic
field.

Although in the case of theories involving parity conserving interactions
the problem can, in principle, be reduced to a question of computing time, in
the case of parity violating interactions the problem is more fundamental
{\bf [1]}.
In fact almost nothing is known on non-perturbative aspects in chiral gauge
theories as compared with the case of vector like theories. The basic
difference is that no regularization which is manifestly gauge invariant is
known (perhaps it does not exist).

When a gauge theory with fermionic matter fields is formulated on a lattice,
then one finds that the only way to translate the gauge invariant solution of
the doubling problem used in the case of vector like theories to the chiral
case requires to use a scalar field. But,in fact, a recent work {\bf [2]} has
shown that these formulations fail because it is not possible to avoid really
the doubling problem and the theories become vector like.

 The other alternative is to use directly the solution of the doubling problem
in the parity conserving case loosing gauge invariance at the level of the
lattice action. It is an open problem to see whether a gauge invariant
continuum limit can be recovered and if this gives a framework were
non-perturbative calculations can be done in practice.

At this point it is natural to look for a different starting point for a
formulation of a fermionic field which could give a new perspective to the
problem of including non-perturbative effects in the parity conserving case,
and  which could also be applied to chiral gauge theories.

In a previous work {\bf [3]} ,a second order formulation of fermions
\footnote{$^\ast$}{
In fact the possibility to consider a second order lattice action for
fermions has been considered previously {\bf [4]} }
, based on the identification of decoupled field components in the usual
Dirac action,  has been introduced. It eliminates half of the integration
variables in the  path integral formulation and it leads to a different
approach to chiral gauge theories. It can be interpreted as an elaborated
version of an attempt by Feynman and Gell-Mann {\bf [5]} to understand the
properties of the weak interactions as a consequence of using a formulation
based on the use of two components spinors instead of Dirac spinors.

The aim of this work is to explore the possibility to define a lattice second
order formulation, first in the case of parity conserving gauge interactions,
and the possible extension to chiral gauge theories.

The starting point of the second order formulation {\bf [3]} is the
identification of the variables
$$\chi_L = {1\over \sqrt{m}} \psi_L, \,\,\,\,\, \chi_R = \sqrt{m} ( \psi_R  +
{\cc{\Dsl}_{L}\over m}\psi_L ),$$
$${\bar \chi_R} = {1\over \sqrt{m}} {\bar \psi_R} , \,\,\,\,\, {\bar \chi_L} =
\sqrt{m} ( {\bar \psi_L}  -  {\bar \psi_R}{ {\Ksl{\Dsl}}_{R}\over m} ),
\eqno(1)$$
where $\psi_L, \psi_R$ are the two chiral components of a Dirac field, which
allow to rewrite the standard (euclidean) gauge invariant Dirac action
$$ S^{(1)} =  \int d^Dx \biggl[ m {\bar \psi}\psi + {1\over 2}{\bar \psi}
( {\cc{\Dsl}} -  {\Ksl{\Dsl}}) \psi \biggr], \eqno(2)$$
into the form
$$S^{(1)} =  \int d^Dx \biggl[ m^2 {\bar \chi}_R{ \chi}_L +
  {\bar \chi}_R ( {\Ksl {\Dsl }_{R}}{\cc{\Dsl }}_{L} )\chi_L +
{\bar \chi}_L \chi_R \biggr].\eqno(3)$$

{}From this expresion one identifies ${\bar \chi}_L, \chi_R$ as auxiliary non-
propagating fields leading to a second order formulation
$$S^{(2)} =  \int d^Dx \biggl[ m^2 {\bar \chi}_R{ \chi}_L +
  {\bar \chi}_R ( {\Ksl {\Dsl }_{R}}{\cc{\Dsl }}_{L} )\chi_L
\biggr],\eqno(4)$$
which reproduces at the perturbative  level all the results of the usual Dirac
formulation including the $U(1)$ anomaly {\bf [3]}.

In the chiral case a
mass term which is  essential for the derivation of the second order
formulation
is not present in the Dirac action, but it is possible, in principle, to start
directly with the second order action $S^{(2)}$ with $m=0$ and the parity non-
invariance of the gauge interaction is reflected on the different action of the
covariant derivative on $\chi_L$ and ${\bar \chi}_R$.

In order  to explore further the possibility to formulate a theory with
fermionic degrees of freedom with a second order action, the next step is to
introduce explicitly a regularization. One way to do that, which can be taken a
a starting  point for a non-perturbative  study, is based on the formulation of
the theory on a discrete lattice of spacetime points (a hypercubic lattice for
example).

A naive discretization of the usual Dirac action leads to the well known
doubling problem (the lattice action describes $2^D$ fermionic degrees of
freedom for each fermionic field). This problem, which technically can be
related to the first order action {\bf [6-7]}, is connected with the $U(1)$
anomaly. Since the anomaly is contained in the second order formulation {\bf
[3]} one can anticipate that the doubling problem will also be present in a
discretization of the second order action although perhaps in a different way.

In order to show that and to study the way to avoid the doubling problem in
the second order lattice formulation, let us translate to a lattice formulation
the steps going from the Dirac action to the second order action.

The starting point is the naive discretization of the free Dirac action
$$S^{(1)}_f = \Sn \biggr[ m\,{\bar \psi}_{\nf} \psi_{\nf} + {1\over 2} {\bar
\psi}_{\nf}  \left( {\cc{\Qsl}}^\pm - {\Ksl {\Qsl}}^{\pm} \right) \psi_{\nf}
\biggr],  \eqno(5)$$
which is just the first order action in (1), particularized to the free case
$(A_\mu = 0)$, when the integral is replaced by a sum over the hypercubic
lattice. The  points of the lattice are characterized by a D-vector ${\bf n}$
whose components are  integers in lattice units (a), the Dirac field  $\psi
(x)$
is replaced by a Grassmann variables $\psi_{\nf}$ at each lattice point and the
derivative is  replaced by the symmetrized finite difference
$$ {( {\Tr}^\pm_\mu \psi )}_{\nf} = {{\psi_{{\nf} + {\hat \mu}}
- \psi_{{\nf} - {\hat \mu}}}\over 2}, \eqno(6)$$
where ${\nf} \pm {\hat \mu}$ is a vector with components
$${({\nf} \pm {\hat \mu})}_\mu = {\nf}_\mu \pm {\bf 1};\,\,\,\,\,
{({\nf} \pm {\hat \mu})}_\nu = {\nf}_\nu\,\,\, for\,\,\, \nu \ne \mu.
\eqno(7)$$

The doubling problem is identified at this level in the discretized free
fermion propagator, which is the free Dirac propagator with the momentum
replaced by
$${\hat P}_\mu = {1\over a} \sin (a\,p_\mu), \eqno(8)$$ vanishing not only at
$p_\mu =0$, but also at $p_\mu = {\pi\over a}$, as a  consequence of the
discretization of the derivative in (6).

Nothing prevents us to repeat step by step the identification of decoupled
variables and the naive second order action on the lattice will be
$$S^{(2)}_f =  \Sn\biggl[  m^2\,{\bar \chi}_{R} \chi_{L} +
{\bar \chi}_R {\Ksl {\Qsl}}^\pm_{R} {\cc{\Qsl}}^\pm_{L} \chi_L \bigg],
\eqno(9)$$ where the dynamical variables ${\bar \chi}_R, \chi_L$, as well as
the decoupled auxiliary fields, will be given by (4) with the derivatives
replaced  by the finite difference (6).

Although the lattice action is of second order, as in the case of a bosonic
field, the doubling problem is present due to the symmetrized finite difference
approach for the derivative. This had to be the case since the action
$S^{(2)}_f$  is nothing but a reformulation of the naive action
$S^{(1)}_f$.

The only difference with the discretization of the Klein-Gordon action of a
free bosonic field without any doubling problem is that in this case the
kinetic operator is given by $\phi^+ \,\,{\Ksl {\Tr}}^+_\mu {\cc{\Tr}}^+_\mu
\,\,\phi$ with the finite difference
$$ {( {\Tr}^+_\mu \phi )}_{\nf} = \phi_{{\nf} + {\hat \mu}}
- \phi_{\nf}, \eqno(10)$$
instead of the symmetrized version in (6).

{}From this comparison it is very easy to identify what are the terms to be
added
to the naive action $S^{(2)}_f$ in order to avoid  the doubling problem. By
using
the discretized version of the identity
$${\dsl}_{R} {\dsl}_{L} = \partial_\mu \partial_\mu, \eqno(11)$$
which is a consequence of the Dirac algebra and the conmutation relation of the
free derivatives, one is lead to consider
$${\bar S}^{(2)}_f = S^{(2)}_f +  \Sn {\bar \chi}_R \left(
{\Ksl {\triangle}}^+_\mu {\cc {\triangle}}^+_\mu -
{\Ksl {\triangle}}^\pm_\mu {\cc {\triangle}}^\pm_\mu \right) \chi_L,
\eqno(12)$$
as the free second order action.

The second term guarantees that the free
propagator will have the same dependence in momentum as the bosonic propagator
on the lattice and then it eliminates the doubling problem. It plays the same
role as the Wilson term in the first order action, being an irrelevant term in
the naive classical continuum limit and giving a mass of the order of the
cutoff to all the additional degrees of freedom which are lattice artifacts
\footnote{$^\ast$}{In fact the operator in the second term
$${\Ksl {\triangle}}^+_\mu {\cc {\triangle}}^{+}_\mu -
{\Ksl {\triangle}^{\pm}}_\mu {\cc {\triangle}}^{\pm}_\mu =
\biggl({{\Ksl {\triangle}}^+_\mu
 - {\Ksl {\triangle}}^-_\mu\over 2}\biggr)\,\,\biggl(
 {{\cc {\triangle}}^+_\mu
 - {\cc {\triangle}}^{-}_{\mu}\over 2}\biggr),$$
is just the square of the operator which appears in the Wilson term.}.

In the presence of the gauge field one has the corresponding discretized
versions of the covariant derivative
$${( D^\pm_\mu \psi)}_{\nf } = {U_{{\nf},{\mu}}
\psi_{{\nf} +  {\hat \mu}} - U_{{\nf},-{\mu}}
\psi_{{\nf} -  {\hat \mu}}\over 2},\eqno(13)$$
$$ {(
D^+_\mu \psi)}_{\nf} = U_{{\nf},\mu} \psi_{{\nf} +  {\hat
\mu}} - \psi_{\nf}, \eqno(14) $$
where $U_{{\nf},\mu}$ is the gauge
variable defined on the link between  the points ${\nf}$ and $ {\nf} +  {\hat
\mu}$. Covariant derivatives have a non trivial conmutator and then one has
$${\Ksl {\Dsl }_{R}}{\cc{\Dsl }}_{L} = {\Ksl {D}}_\mu\,{\cc
 D}_\mu  + {1\over 4} {[\gamma_\mu, \gamma_\nu]}_{RL} ({\Ksl
{D}}_\mu\,{\cc D}_\nu -  {\Ksl {D}}_\nu\,{\cc D}_\mu ),
\eqno(15)$$
where the first term is the covariant version of the free kinetic
operator and the second term is due to the gauge interaction.

In order to
identify the second order lattice formulation of a gauge theory what one has to
do is to combine  the free lattice action ${\bar S}^{(2)}_f$, when the
derivatives $\triangle_\mu$ are replaced by covariant derivatives, with an
appropiate discretization of the contribution involving the conmutator of
covariant derivatives.

The simplest action one finds following this procedure is
$$\eqalignno{S^{(2)}_{lat} = &\Sn \biggl[ m^2 {\bar \chi}_R \chi_L +
{\bar \chi}_R {\Ksl { D}}_\mu^+ {\cc {D}}_\mu^+\chi_L
\biggr] +\cr &  +{1\over 4} \Sn {\bar \chi}_R {[\gamma_\mu, \gamma_\nu]}_{RL}
({\Ksl {D}}^\pm_\mu\,{\cc D}^\pm_\nu -
{\Ksl {D}}^\pm_\nu\,{\cc D}^\pm_\mu )
\chi_L , &(16) \cr}$$
which corresponds to the minimal modification of the
naive discretization of the Dirac action required in order to avoid the
doubling problem. The term involving the conmutator of covariant derivatives
does not require any modification and it is taken from the naive Dirac action
which involves the symmetrized discrete version of the covariant derivative
(13).

The action $S^{(2)}_{lat}$ is obtained from the naive  discretization of the
Dirac action by adding the term
$$\Delta S_{lat} =  \Sn {\bar \chi}_{R} \left( {\Ksl { D}}_\mu^+ {\cc
{D}}_\mu^+  - {\Ksl { D}}_\mu^\pm {\cc {D}}_\mu^\pm \right) \chi_L, \eqno(17)$$
which is the gauge invariant version of the term identified in the free case as
a way to solve the doubling problem.

The action $S^{(2)}_{lat}$ can be taken as a spacetime lattice regularization
of the second order formulation based on the action $S^{(2)}$. As it is always
the case in  a lattice regularization, manifest covariance is lost, and all one
has is an invariance  under the discrete transformations leaving invariant the
spacetime lattice (${\pi\over 2}$ rotations in the hypercubic lattice) which
one expects is enough to recover a covariant continuum limit.

The general analysis of the anomaly in the second order formulation {\bf [3]}
can be applied also to the lattice regularization. It was based on the
identification  of a regularization independent part and a combination of
covariance arguments, to fix the more general tensor structure in the quadratic
contribution to the effective action, together with the constraints imposed by
gauge invariance. It is very easy to see that the invariance under lattice
rotations is enough to fix the same tensor structure. Then a lattice
regularization which is manifestly gauge invariant and which avoids the
doubling
problem, as it is the case for
$S^{(2)}_{lat}$ ,allows to reproduce the standard result for the
$U(1)$ anomaly. A part of the anomaly comes from the non invariance  of the
naive discretization of the second order action under a chiral transformation
of the dynamical degrees of freedom $\chi_L, {\bar \chi}_R$, and the rest
of the anomaly comes from  the fermionic expectation value of the variation of
the term $\Delta S_{lat}$ which is  required in order to solve the doubling
problem.

Once more one recognizes in the second order formulation the connection
between the doubling problem and the anomaly. The only difference with the
analysis based  on the Dirac action is that the elimination of the decoupled
fields $\chi_R, {\bar \chi}_L$ already introduces a violation of chiral
invariance before the regularization is introduced.

The action $S_{lat}^{(2)}$ is just one of the possible ways to discretize the
second order action $S^{(2)}$. It is natural to expect such an ambiguity for
any regularization. The only properties that the correction
$\Delta S_{lat}$ eliminating the doublers of the naive discretization should
satisfy in order to reproduce the anomaly are gauge  invariance and lattice
rotational invariance. Then, unless unitarity imposes some additional
constraints ( we have not found a practical method to study this question), any
lattice regularization  free of doubling which differs from $S_{lat}^{(2)}$ in
gauge invariant and lattice rotational invariant terms which are irrelevant in
the naive continuum limit, can be used as a lattice second order formulation
of a gauge theory.

At the perturbative level, a lattice second order formulation is equivalent to
the standard first order formulation. From a practical point of view, as
a consequence of the identification of the decoupled auxiliary fields
$\chi_R, {\bar \chi}_L$, one has a path integral formulation with half of
the anticonmuting variables to integrate for a given lattice which, in
principle, can be associated to an important \lq \lq saving" of computing
time.

This point, together with some other possibles differences at the
non-perturbative level, deserve further investigation. In particular the
approach to the chiral massless limit could be one of the problems where the
second order formulation could bring some new perspectives as a consequence of
the different realization of chiral symmetry.

A related and still more promising possible application of a second order
lattice formulation is to try to extend the action $S_{lat}^{(2)}$ to a chiral
gauge theory. In fact a non-perturbative formulation of a chiral
gauge theory is still an open problem; the attempts to extend a lattice first
order formulation to the chiral case presents some basic difficulties which
still do not have a satisfactory solution. This makes interesting any
possibility  to approach this problem from a different point of view, like the
second order formulation.

One way to approach this problem, which is similar to the approach followed in
the first order formulation {\bf [1]} ,is based on the introduction of a scalar
field coupled to the fermion field in order to give a mass to the fermion in a
gauge invariant way. One can also apply this coupling to
give masses  to the replica fermions of the order of the cutoff in the
chiral case through a gauge invariant Wilson-Yukawa term in the lattice action.

The translation of this
approach to the second order formulation takes as starting point an action
$$S^{(1)}_{ch} =  \int d^Dx\,\, \biggl[ {\bar \psi}_R \phi \psi_L +
{\bar \psi}_L \phi^\ast \psi_R +
{1\over 2}{\bar \psi}( {\cc{\Dsl}} -  {\Ksl{\Dsl}}) \psi \biggr], \eqno(18)$$
where the mass term is replaced by a coupling to the scalar field
$$\eqalignno{&\chi_R = \psi_R + {1\over \phi^\ast} {\cc {\Dsl}}_{L} \psi_L ,
\cr & {\bar \chi}_L = {\bar \psi} _L -
{\bar \psi}_R {\Ksl {\Dsl}}_{R}{1\over \phi^\ast}, &(19) \cr}$$
and then one has
$$S^{(1)}_{ch} =  \int d^Dx\,\, \biggl[\,\, {\bar \psi}_R \phi \psi_L +
{\bar \psi}_L{\Ksl {\Dsl }}_{R} {1\over \phi^\ast}
{\cc {\Dsl}}_{L} \psi_L + {\bar \chi}_L \phi^\ast \chi_R \biggr], \eqno(20)$$

The only term involving the variable $\chi_R$ is a trivial factor which can be
absorved into a redefinition of the fermionic measure, which leads to
$$S^{(2)}_{ch} =  \int d^Dx\,\, \biggl[\,\,\, {\bar \psi}_R \phi \psi_L +
{\bar \psi}_L{\Ksl {\Dsl }}_{R} {1\over \phi^\ast}
{\cc {\Dsl}}_{L} \psi_L \biggr]. \eqno(21)$$

It is  not clear how to proceed from this non-polynomic action. If one formally
translates the lattice regularization of the parity conserving case to the
chiral action then one would have
$$\eqalignno{& S^{(2)}_{ch} = \Sn \,\, \biggl[\,\,\,{\bar \psi}_R \phi \psi_L +
{\bar \psi}_R {\Ksl { D}}^+_\mu {1\over \phi^\ast} {\cc {D}}^+_\mu \psi_L
\,\, \biggr] \,\, + \cr & + {1\over 4} \Sn {\bar \psi}_R {[\gamma_\mu,
\gamma_\nu]}_{RL}  \left({\Ksl {D}}^\pm_\mu\,{1\over \phi^\ast}\,{\cc
D}^\pm_\nu
-  {\Ksl {D}}^\pm_\nu\,{1\over \phi^\ast}\,{\cc D}^\pm_\mu \right)
\psi_L, &(22) \cr}$$
which is the translation to the second order formalism of the formulation of a
chiral theory based on the introduction of a scalar field. It is difficult to
imagine why the difficulties which appear in the first order formalism
{\bf [2]} should not be present also in this case.

Another approach to the problem of regularizing a second order formulation of a
chiral gauge theory is based on a direct discretization of the action (4)
with $m=0$ and the covariant derivative acting differently on ${\bar \chi}_R,
\chi_L$.  In this case one has the gauge variables $U_{{\nf},\mu}$ and
${\bar U}_{{\nf},\mu}$ which correspond to the group element in the
representation of $\chi_L$ and ${\bar \chi}_R$ respectively.

In the abelian case one has
$$ U_{{\nf}, \mu} = \exp \left( ieA_{{\nf}, \mu}\right);\,\,\,\,\,
{\bar U}_{{\nf},\mu} = \exp \left( -i{\bar e}A_{{\nf}, \mu}\right). \eqno(23)$$

The symmetrized discretized covariant derivatives will be
$${({\cc {D}}^\pm_\mu \chi_L)}_{\nf } = {U_{{\nf}, \mu}
\chi_{L, {\nf} +  {\hat \mu}} - U_{{\nf}, -\mu}
\chi_{L, {{\nf}- {\hat \mu}}}\over 2},\eqno(24a)$$
$${(\bar \chi}_R {\Ksl { D}}^\pm_\mu )_{\nf } =
{{\bar \chi}_{R, {\nf} +  {\hat \mu}} {\bar U}_{{\nf}, \mu} -
{\bar \chi}_{R, {{\nf}- {\hat \mu}}} {\bar U}_{{\nf}, -\mu}\over 2}
\eqno(24b)$$ and
 $$ {({\cc
{D}}^+_\mu \chi_L)}_{\nf} = U_{{\nf},\mu} \chi_{L,{{\nf}} +  {\hat
\mu}} - \chi_{L,{\nf}} \,\,\, , \eqno(25a) $$
$$ {({\bar \chi}_R {\Ksl {D}}^+_\mu )}_{\nf} =
{\bar \chi}_{R,{\nf} +  {\hat
\mu}}{\bar U}_{{\nf},\mu}-{\bar\chi}_{R,{\nf}} . \eqno(25b) $$

Then one can translate the second order lattice formulation in the parity
conserving case to the chiral case and one has
$$\eqalignno{ {\bar S^{(2)}}_{ch}  = & \Sn
{\bar \chi}_R {\Ksl { D}}^+_\mu {\cc D}^+_\mu \chi_L \,\,\, + \cr &
+{1\over 4} \Sn {\bar \chi}_R {[\gamma_\mu, \gamma_\nu]}_{RL}
({\Ksl {D}}^\pm_\mu {\cc D}^\pm_\nu -
{\Ksl {D}}^\pm_\nu\,{\cc D}^\pm_\mu )
\chi_L, &(26) \cr}$$
as a candidate for a lattice formulation of a chiral gauge theory.

The possibility to define  a continuum limit which corresponds to a chiral
gauge invariant relativistic quantum field theory  from this lattice action
remains as an open question. Note that the lattice action is not gauge
invariant
because it is already a regularization of a gauge non invariant second order
action, and the additional term required to eliminate the doubling problem
violates also gauge invariance. One can expect that the conservation of the
current coupled to the gauge field in the case of an anomaly free fermion
content will allow to find a gauge invariant continuum limit after appropiate
counterms are added. This seems to be a necessary step in order to get a
chiral gauge invariant theory also in the approach based on the first order
formalism.

Another non trivial question is to see whether
the continuum limit corresponds to a unitary theory, also as a consequence of
the conservation of the current coupled to the gauge field. We do not known how
prove the validity of the formulation of a chiral gauge theory based on the
lattice action ${\bar S}^{(2)}_{ch}$. As a first step, a perturbative and
non-perturbative  analysis of two dimensional models is under
investigation.

Other open problems, like the introduction of four fermion interactions in the
second order formalism, the study of spontaneous symmetry breaking and
dynamical generation of mass, will be the subject of a future work.

To summarize a second order formulation based on the identification of a
combination of fermionic field components with no dynamics has been proposed as
a way to study a gauge invariant parity conserving theory. The possibility to
apply this formalism to the case of chiral gauge theories has been also pointed
out with the new perpectives that it can open on the dynamics of these
theories.

A lattice regularization of the second order parity invariant formulation free
of the doubling problem has been presented. The identification of the decoupled
auxiliary fields leads to a path integral formulation with half of the
integration variables which can simplify considerably any numerical study.

An attempt to give a non-perturbative formulation of chiral gauge theories on
the lattice has been presented in detail. It remains for the future to test the
validity of this formulation and to apply it to study some non-pertubative
effects of the standard model.

We would like to thank H.B. Nielsen by useful discussions.
\vskip 0.5cm
\centerline{\bf References}
\vskip 0.25cm
\item{\bf[1]} For a review and references see M.F.L.Golterman, Nucl.Phys. {\bf
20B } (Proc.Suppl.) (1991)528.
\item{\bf[2]} M.F.L.Golterman, D.N.Petcher and J.Smit, Nucl.Phys. {\bf
370B}(1992)51.
\vfill \eject
\item{\bf[3]} J.L. Cort\'es, J. Gamboa and L.Vel\'azquez, {\it Second Order
Formalism for Fermions} Zaragoza-Orsay Preprint, DFTUZ 92/13.
\item{\bf[4]} T. Banks and A. Casher, Nucl. Phys. {\bf 169B}(1980)103;
A.C. Longhitano and B. Svetitsky, Phys Lett. {\bf 126B}(1983)259.
\item{\bf[5]} R.P. Feynman and M. Gell-Mann, Phys. Rev. {\bf 109}(1958)193.
\item{\bf[6]} K.G. Wilson e.g  {\it
New Phenomena in Subnuclear Physics}, ed.  A. Zichichi, Plenum (1977).
\item{\bf[7]} H.B. Nielsen and M. Ninomiya, Nucl. Phys. {\bf 185B}(1981)20;
{\bf 193B}(1981)173; Phys. Lett. {\bf 105B}(1981)219.

\end